\documentclass[runningheads]{llncs}
\usepackage[T1]{fontenc}
\usepackage{graphicx}
\begin{document}
\title{RawNet: Fast End-to-End Neural Vocoder}
\author{Yunchao He \and
Yujun Wang
}
\authorrunning{Y. He et al.}
\institute{Xiaomi Corporation, Beijing, China \\
\email{heyunchao@xiaomi.com} \email{wangyujun@xiaomi.com}\\
\url{https://www.mi.com/}}
\maketitle              % typeset the header of the contribution
\begin{abstract}
Neural network-based vocoders have recently demonstrated the powerful ability to synthesize high-quality speech. These models usually generate samples by conditioning on spectral features, such as Mel-spectrogram and fundamental frequency, which is crucial to speech synthesis. However, the feature extraction procession tends to depend heavily on human knowledge resulting in a less expressive description of the origin audio. In this work, we proposed RawNet, a complete end-to-end neural vocoder following the auto-encoder structure for speaker-dependent and -independent speech synthesis. It automatically learns to extract features and recover audio using neural networks, which include a coder network to capture a higher representation of the input audio and an autoregressive voder network to restore the audio in a sample-by-sample manner. The coder and voder are jointly trained directly on the raw waveform without any human-designed features. The experimental results show that RawNet achieves a better speech quality using a simplified model architecture and obtains a faster speech generation speed at the inference stage.

\end{abstract}
\section{Introduction}

In speech synthesis, the vocoder is essential in extracting acoustic features and recovering speech. Although neural networks have mainly carried out the recent process of restoring speech, feature extraction still relies heavily on manual-designed steps.

Traditional vocoding approaches \cite{world} \cite{straight} \cite{Vocaine} are commonly composed of a speech analysis module and a waveform generation module. The analysis module is responsible for extracting the acoustic features from the raw waveform, while the waveform generator reconstructs the audio signal from the features. In the speech synthesis task, the commonly used acoustic features are extracted based on complicated human-designed speech production procedures, such as the source-filter model \cite{sf1} \cite{sf2} \cite{sf3}. In \cite{world} and \cite{straight}, the acoustic features include the log fundamental frequency (lf0), voice/unvoiced binary value (UV), the spectrum and band aperiodicities. However, the underlying assumption of these models makes it complicated to generate the waveform and often introduces some flaws and artifacts into the generated speech.

In addition to the traditional human-designed feature extraction, the raw waveform-based methods have been explored in many speech-related tasks. In speech recognition, \cite{zeghidour2018end} proposed the recognition model based directly on the raw waveform and achieves a better result than the model trained with hand-crafted acoustic features. In \cite{raw_fft_sincnet}, the raw waveform is directly fed into the neural model for both speech and speaker recognition tasks. \cite{raw_fft_sincnet} shows the benefits of model convergence, performance, and computational efficiency. \cite{soundnet} explores the representative feature directly from many sound data and yields the state-of-art result in the acoustic object classification task. In \cite{fu2017raw}, a fully convolutional network is used to enhance speech directly using raw waveform as model input and target. 

Another important function of the vocoder is to restore audio from features. Recently, neural vocoders use neural networks to directly learn the transformation from the acoustic features to audio waveform such as WaveNet \cite{oord2016wavenet}, LPCNet \cite{LPCNet}, WaveGlow \cite{prenger2018waveglow}, HiFi-GAN \cite{hifigan}, MelGAN \cite{melgan} and FFTNet \cite{fftnet}. They greatly improve the quality of speech synthesis compared with the traditional methods by eliminating the complicated human-designed speech generation steps. However, the waveform generation is slow due to the complicated model structure. In addition, the performance of these neural vocoders is also affected by the conditioned acoustic features.

As the acoustic features bridge the gap between the acoustic model and the vocoder, how to choose appropriate features is essential. A text-to-speech model often uses a low-dimensional representation of the raw waveform, often predicted by the acoustic models, then used by the vocoder to reconstruct the predicted waveform. As the acoustic model is trained to minimize the gap between the ground-true acoustic features and predicted features, this paper obtains acoustic features taking the following three factors into account: 1) if it's easy to be predicted by the acoustic model, 2) if it could represent the raw waveform expressively and compactly, and 3) if it is able to reconstruct waveform with high quality. Based on the three factors, we evaluate if it's an appropriate acoustic feature for speech synthesis.

Inspired by the success of the neural network-based methods, it is possible to further improve the existing neural vocoder by embedding the feature extractor model as part of the vocoder network and jointly optimizing the whole vocoder framework. In this paper, we propose an entire end-to-end neural vocoder architecture called \emph{RawNet} leveraging the powerful ability of the neural network to extract features and restore audio. The term \emph{end-to-end} aims to emphasize that RawNet directly takes the raw signal as input for feature extraction and generates raw waveform as output. It is similar to an auto-encoder model but considers extracted features' predictability. RawNet comprises a coder network responsible for capturing acoustic features from the raw waveform and a voder network for reconstructing high-quality speech waveform. These two components correspond to a traditional vocoder's analysis and synthesis module. 

The rest of the paper is organized as follows. Section 2 introduces some related works, including speech feature extraction, the application of the auto-encoder model for processing speech signals, and some popular neural vocoders in the field of speech synthesis. Section 3 presents the proposed model RawNet and some crucial training strategies. Section 4 shows the experimental settings and results. Conclusion and future works are provided in section 5.

\section{Related Works}

There is some research on employing an auto-encoder for extracting relative parameters for speech synthesis tasks. \cite{vishnubhotla2010autoencoder} \cite{raitio2014deep} use an autoencoder to extract excitation parameters, which is required by a traditional vocoder. In \cite{cstr_deep_autoencoder}, an autoencoder-based, non-linear, and data-driven method is used to extract low-dimensional features from the FFT spectral envelope instead of using the speech analysis module based on human knowledge. \cite{cstr_deep_autoencoder} also concludes that the proposed model outperforms the one based on the conventional feature extraction. The difference between RawNet and the methods mentioned above is that RawNet directly takes waveform samples as input instead of treating the autoencoder as a feature dimension reduction method.

In addition, the recently emerging jointly training method widely adopts the modeling technique on the original waveform in speech synthesis. The VITS \cite{vits} model embeds a parallel waveGAN vocoder in the end-to-end speech generation model, in which a hidden representation z of the waveform is learned. SANE-TTS \cite{sanetts} extends the VITS model to multilingual text-to-speech by disentangling speaker and language information from text encoding. Different from these methods, RawNet is trained independently to acoustic models.

Our work differs from these in that we use an auto-encoder-based model framework for extracting higher representative features for speech synthesis. The novelties and contributions of our work are that: 1) We directly extract the desired features from the raw waveform instead of modeling on the FFT spectral envelope, 2) We embed the feature extraction network into the unified end-to-end vocoder model rather than using human-designed acoustic features, 3) The learned acoustic feature is independent of the acoustic models.

\begin{figure}[htb]
\begin{center}
\includegraphics[width=0.6\textwidth]{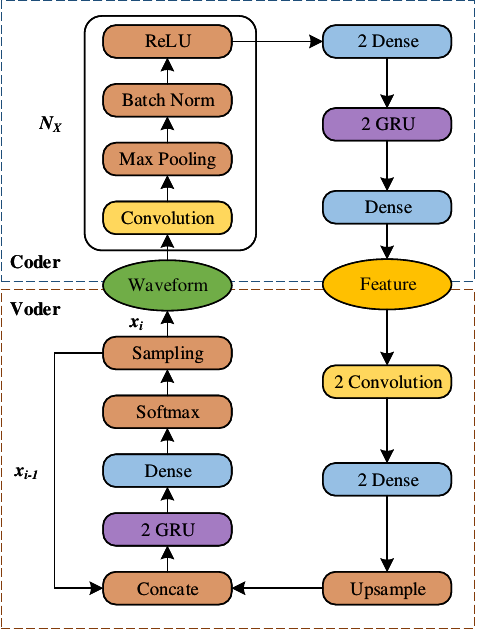}
\end{center}
\caption{The model architecture of RawNet mainly consists of two parts: coder and voder. The upper part is the coder network which extracts acoustic features from raw waveform, and the bottom part is the voder network which generates speech from learned features.} \label{fig:rawnet_fig}
\end{figure}

\section{RawNet}

This section introduces the RawNet model. Figure~\ref{fig:rawnet_fig} shows its overview architecture, which includes a coder network that extracts acoustic features from the raw waveform and a voder network that generates waveform conditioned on the learned acoustic features. These two parts are jointly trained in a single model. Still, they could be used separately in text-to-speech tasks, corresponding to a traditional vocoder system's analysis and synthesis procedure. More details are provided in this section. 

\subsection{Coder network: Automatic Feature Extraction}

The coder network is used for automatic feature extraction from the raw waveform. The main components of the network are stacked convolutional layers, dense layers, and GRU layers, as shown in the upper part of figure ~\ref{fig:rawnet_fig}. The coder network learns the high-level audio representation through a series of lower-level filters by stacking multiple convolutional layers. The convolutional layer is similar to the one proposed in \cite{soundnet}, which is used to learn the sound representation. To better preserve the time-serious nature of the extracted features, we replaced them with causal convolutional layers. By extending the network with GRU and dense layers, we empower the model with the ability to capture the long-term relationship.

Given different audios varying in temporal length, the coder network is expected to convert the samples in time domains into frame sequences in the frequent-like space. To get the desired frame number and window size, we control the stride step in the convolutional layers and the pooling size of the pooling layers. As convolutional layers are invariant to location, we convolve multiple layers to control the output length. Consequently, the learned acoustic features' frame size is only determined by the convolutional and pooling layers.

\subsection{Voder network: Restoring audio}

The voder network is for restoring audio from the acoustic features either predicted by acoustic models or directly learned by the coder network. Its structure is similar to that of LPCNet \cite{LPCNet}, but we make some modifications. In LPCNet, the current predicted sample and excitation, global features from the frame-rate network, and linear prediction are used as inputs to generate the following samples. Unlike LPCNet's complicated input information, the voder network of RawNet only takes the current predicted sample and the conditioning acoustic features as input which could simplify the network and reduce inference complexity. The concatenated inputs are fed to the subsequent layers to predict the next sample. 

The conditioned acoustic features are first fed into two convolutional layers, followed by two dense layers. The output of the dense layer is in frame length and then is up-sampled to the sample length. To speed up the audio generation process, we apply a simple up-sampling method by repeating the inputs K times, where K is the frame size determined by the Coder network. The up-sampled features, concatenated with the previously predicted sample, are fed into 2 GRU layers, followed by a DualFC layer and a softmax function. The output samples are generated in a sample-by-sample manner for better speech quality.

To normalize the value of input samples and make the Voder network more robust on prediction noise, we apply the \(\mu\)-law algorithm to companding transform the 16-bit samples to an 8-bit discrete representation. An embedding representation is learned for each \(\mu\)-law level, essentially learning a set of non-linear functions applied on the \(\mu\)-law values.

\subsection{Sampling method}

It's reported in LPCNet and FFTNet that directly sampling from the output distribution can sometimes result in excessive noise. FFTNet proposed a conditional sampling method to address this problem, multiplying the output logits by a constant value, i.e., \emph{c=2}, for voiced sounds and remaining for the unvoiced region. LPCNet replaces the binary voicing decision with a pitch correlation, which could be used to scale the output logits continuously.

We compare multiple sampling strategies, including multinomial sampling, conditional sampling, LPCNet's pitch correlation-based sampling, and the simple argmax method, to get a better sampling method. We find that the simple argmax method generates clear samples with the least noise, which is in accord with the results of the original WaveNet \cite{oord2016wavenet} and \cite{oord2016pixel}. One possible explanation is that the learned acoustic feature in the Coder network is helpful for sample prediction even though the unvoiced region is difficult to reconstruct. 

In the conditional and LPCNet's sampling method,  pitch and pitch correlation are required to scale the output logits, while the proposed Coder network does not learn this information explicitly. For comparison, we extract pitch and pitch correlation as additional acoustic features. The REAPER \cite{reaper} tool extracts these features in the comparison experiment.

\subsection{Noise injection}

The noise injection strategy is adopted to ensure that the model sees different training data at each training iteration and avoids over-fitting.

Since the synthesized samples inevitably contain noise due to the training error, the generated samples get noisier over time without denoising methods because of the auto-regressive property. We inject Gaussian noise into the input during training in the Voder network to address this problem. The Gaussian noise is sampled from $\mathcal{N}(0, 1)$ distribution and weighted by a factor $\sigma$ to control the noise temperature, gradually increasing from 0 to 0.2. Besides that, Gaussian noises are also injected into the coder network's input.

\subsection{Post-synthesis denoising}

Even though injecting noise enables the networks to see more training data and avoid over-fitting problems, it also introduces a small amount of buzz noise to the silent part of the unvoiced sound. The noise is sometimes audible with a low magnitude and only occurs in the silence part. Therefore, we apply a simple energy-based method \cite{vad}, which is a baseline method in voice activity detection to reduce these noises. Experiments show that this method could almost eliminate these noises. 

\section{Experiments}

To evaluate the power of RawNet, We conduct an AB preference test to compare the quality of the generated speech from RawNet and LPCNet.

\subsection{Experimental setup}

The proposed system can be either speaker-independent or speaker-dependent. We evaluate the model in both settings using three different datasets. The CMU Arctic dataset \cite{cmu_arctic_data} is used to train a speaker-independent vocoder. The CMU ARCTIC consists of around 1150 utterances for each speaker, including females and males. To reduce the accent variance, we select four speakers as training data consisting of two male speakers, \emph{bdl} and \emph{rms}, and two female speakers, \emph{slt} and \emph{clb}. We use a private Chinese dataset MuFei, and a public LJ-Speech 1.1 \cite{ljspeech17} for speaker-dependent experiments. The former contains 20-hour audio from a single female speaker, while the latter consists of about 24-hour audio from a single female speaker. We randomly excluded 1000 samples from each dataset as the test set.

At the training stage, the input of the Coder network is a short audio clip,  which contains 3200 samples (i.e. 200ms for 16k speech). The clip is randomly selected from the original wave. Its output is 20 frames of features, with 64 dimensions per frame. The training epoch in our experiments is 1500, with a batch size of 128*4. The model is trained on four Nvidia P40 GPUs with 22GB memory size. The cross-entropy loss is used as the loss function in the experiments. The weight matrices of the network are initialized with normalized initialization, and the bias vectors are initialized to be 0. AMSGra \cite{adam} ] optimization method (Adam variant) is used to update the training parameters with an initialized learning rate of 1e-2.

\subsection{Subjective evaluation}

\begin{figure}[htb]
\begin{center}
\includegraphics[width=0.9\textwidth]{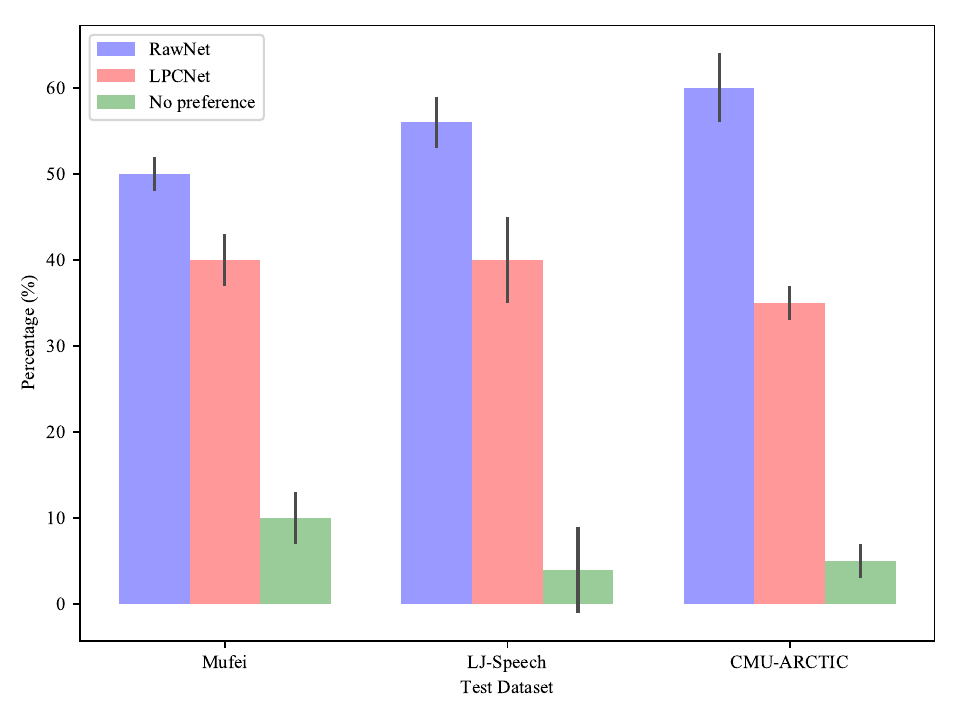}
\end{center}
\caption{A/B Preference Test Result of RawNet and LPCNet of three different datasets.} \label{fig:abtest}
\end{figure}

AB preference tests were conducted to assess the generated speech quality. In AB preference tests, for each task, we randomly selected 15 paired samples A and B from RawNet and LPCNet. There were 20 raters participating in the evaluation, with ten female and ten male raters. The raters were asked to choose the sample with better quality. 

As shown in figure ~\ref{fig:abtest}, the generated speech by RawNet gets more preferences than those of LPCNet. Especially, RawNet has more advantages than LPCNet in the speaker-independent application.

\subsection{Visualization}

\begin{figure}[htb]
\begin{center}
\includegraphics[width=0.7\textwidth]{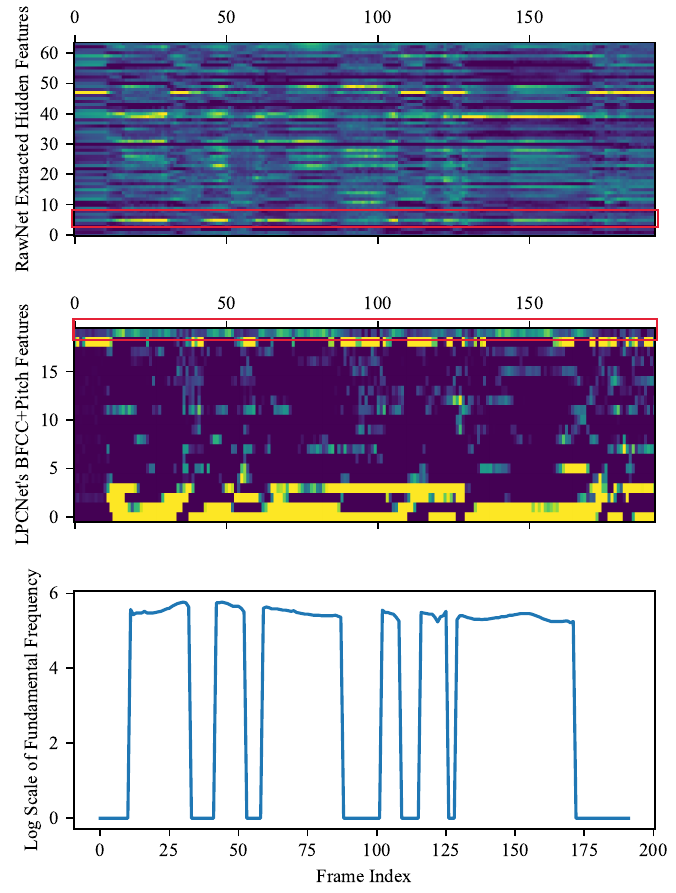}
\end{center}
\caption{These three subfigures compare different acoustic features extracted from the same audio. The top figure shows the features learned from the Coder network. The central figure illustrates BFCC and pitch parameters (colored in red). The bottom figure is the F0 contour.} \label{fig:spectrogram}
\end{figure}
 
Figure ~\ref{fig:spectrogram} illustrates the features extracted by RawNet coder, Bark-frequency cepstral coefficients (BFCC) ~\cite{bfcc} and pitch parameters (period, correlation) used in LPCNet, and the pitch contour extracted from the same audio.
By comparing the region in the red box, we find that the Coder network automatically captures some interpretable features, like pitch information. It indicates that the Coder can learn high-level information from signals without prior knowledge. 

\begin{figure}[htb]
\begin{center}
\includegraphics[width=0.7\textwidth]{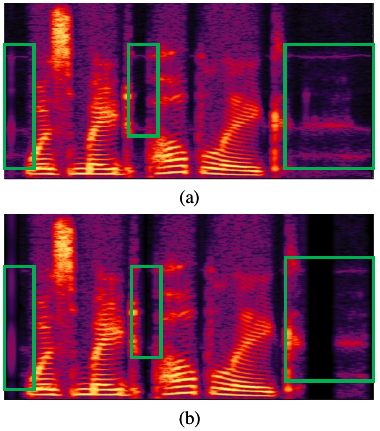}
\end{center}
\caption{This figure compares the spectrogram of generated speech by RawNet with (bottom) and without (top) post-synthesis denoising. The green box region points out the denoising effects.} \label{fig:spec_com}
\end{figure}

The effect of post-synthesis denoising is illustrated in figure ~\ref{fig:spec_com}. After using the post-synthesis denoising strategy, the "click" noise is almost completely removed. 

\section{Conclusion}

This paper proposes a new vocoder that uses a Coder network to learn the representation from the raw waveform and applies a Voder network to restore the waveform. The Coder and voder can be trained jointly. The subjective evaluation shows that our proposed model can produce more natural/preferred speech than the recently proposed LPCNet. Visualization of the learned features helps illustrate that RawNet can extract reasonably meaningful features from the raw waveform. 

\section{ACKNOWLEDGMENTS}

The AB preference test is conducted with the help of the Xiaomi AI Lab PM team. The computation resource is provided and maintained by Xiaomi SRE Team. Xiaomi AI Lab Speech Team provides the private Mufei dataset. We thank them all.

\bibliographystyle{splncs04}
\bibliography{mybib}
\end{document}